\documentclass[iop]{emulateapj}

\usepackage{amsmath}
\usepackage{graphicx}
\usepackage{natbib}
\newcommand{\chandra}{{\it Chandra }}
\newcommand{\xmm}{{XMM-{\it Newton} }}

\shorttitle{Probing X-ray Absorption and Optical Extinction in the ISM Using SNRs}

\begin{document}

\title{Probing X-ray Absorption and Optical Extinction in the Interstellar Medium Using Chandra Observations 
of Supernova Remnants}
\author{Dillon R. Foight{$^{1}$}, Tolga G\"uver{$^{2}$}, Feryal \"Ozel{$^{3}$}, Patrick O. Slane{$^{1}$}}

\affil{$^{1}$ Smithsonian Astrophysical Observatory, 60 Garden Street, Cambridge, MA 02138}

\affil{$^{2}$ Istanbul University, Science Faculty, Department of
Astronomy and Space Sciences, Beyaz{\i}t, 34119, Istanbul, Turkey}

\affil{$^{3}$  Department of Astronomy,  University of  Arizona, 933
    N. Cherry Ave., Tucson, AZ 85721}

\begin{abstract} 
We present a comprehensive study of interstellar X-ray extinction
using the extensive Chandra supernova remnant archive and use our
results to refine the empirical relation between the hydrogen column
density and optical extinction. In our analysis, we make use of the
large, uniform data sample to assess various systematic uncertainties
in the measurement of the interstellar X-ray absorption. Specifically,
we address systematic uncertainties that originate from {\it (i)} the
emission models used to fit supernova remnant spectra, {\it (ii)} the
spatial variations within individual remnants, {\it (iii)} the physical
conditions of the remnant such as composition, temperature, and
non-equilibrium regions, and {\it (iv)} the model used for the
absorption of X-rays in the interstellar medium. Using a Bayesian
framework to quantify these systematic uncertainties, and combining
the resulting hydrogen column density measurements with the
measurements of optical extinction toward the same remnants, we find
the empirical relation $N_{\rm H}\ =
(2.87\pm0.12)\times10^{21}\ A_{\rm V} {\rm\ cm}^{-2}$, which is
significantly higher than the previous measurements.
\end{abstract}


\section{Introduction}

The linear relationship between optical extinction ($A_{\rm V}$) and
hydrogen column density ($N_{\rm H}$) has long been observed and
utilized to estimate the X-ray or optical brightness for new sources
or to fit the broadband spectrum of X-ray sources. It is also used to
obtain distance estimates for X-ray sources using their measured
column densities (see e.g., \citealt{durant2006, guver2010, ratti2010,
  soria2012, neilsen2012}).

The photoelectric absorption by interstellar material causes rapid
attenuation of soft X-rays in the spectra of Galactic
sources. Measuring the extent of this attenuation yields information
about the total column density along the line of sight to the
source. Although this is commonly expressed in terms of the equivalent
hydrogen column density $N_{\rm H}$, in the soft X-ray band ($0.1-10$
keV), it is predominantly caused by abundant heavier elements such as
O, Ne, Si, Mg and Fe. Optical extinction is caused by grains of the
same elements. Because dust tends to follow the metal
  distribution in the ISM, when averaging over many different lines of
  sight and over distances that are larger than the clumping scale of
  the ISM, it is reasonable to assume an approximate linear
  relationship between between $A_{\rm V}$ and $N_{\rm H}$. 

There have been a number of different studies that have sought to
accurately determine the relation between optical extinction and the
hydrogen column density empirically. The methods vary across these
studies, and the results show discrepancies greater than the
statistical errors for each. \citet{reina1973} used X-ray binaries and
extended sources and found a relation of $N_{\rm H} = 1.85 \times
10^{21}\ \times A_{\rm V} {\rm\ cm}^{-2}$ (hereafter, $N_{\rm H}$ is
in units of cm$^{-2}$ and $A_{\rm V}$ is in magnitudes);
\citet{gorenstein1975} used supernova remnants (SNRs) to find $N_{\rm
  H} = (2.22 \pm 0.14) \times 10^{21} \times\ A_{\rm V}$; while
\citet{predehl1995} used a combination of ROSAT point sources and SNRs
and measured $N_{\rm H} = (1.79 \pm 0.03) \times 10^{21}
\times\ A_{\rm V}$.  Recently, \citet{go2009} collected a sample of 22
SNRs, for which the hydrogen column density and optical extinction
were previously measured, and found $N_{\rm H} = (2.21 \pm 0.09)
\times 10^{21} \times A_{\rm V}$. This had the advantage of using the
high quality data from modern X-ray telescopes such as \chandra and
\xmm and focusing on sources with little-to-no intrinsic
absorption\footnote{To illustrate this, consider a remnant
    with a 15 pc radius that has swept up 100 $M_{\mathrm{sun}}$ of
    ISM material (which is a large estimate for most SNRs). The column
    density through the remnant is approximately
    $2.6\times10^{19}~cm^{-3}$, which is at least a factor of ten
    smaller than the $N_{\mathrm{H}}$ values presented in this
    paper.}, but had the disadvantage of being unable to account for
systematic errors that may vary for each published value. For example,
when using values from the literature, there is no way to account for
the variety of choices that are made during the data processing
pipeline. The selected regions or the particular emission models may
be interesting for the objectives of a particular study, but less
ideal for the determination of the hydrogen column density and the
comparison to the optical extinction measurements to determine the
slope of the relation.

\begin{table*}
\centering
\caption{Recovered $N_{\rm H}$ values from simulated data}
\begin{tabular}{ c c c c c}
\hline
\hline
Assumed $N_{\rm H}$ & $T_s$ & $T_{es}$ & $\tau_0$ & Fit $N_{\rm H}$ \\
($10^{22}\ {\rm cm}^{-2}$) & (keV) & (keV) & ($10^{11}$) & ($10^{22}\ {\rm cm}^{-2}$) \\
\hline
\hline
0.5 & 0.3 & 0.15 & 10 &  0.456 $\pm$ 0.094 \\
    &       &         &  5 &  0.412 $\pm$ 0.135 \\
    &       &         &  1 &  0.469 $\pm$ 0.052 \\
    &  0.6 & 0.3  &  10 &  0.525 $\pm$ 0.030 \\
    &       &         &  5 &  0.458 $\pm$ 0.023 \\
    &       &         &  1 &  0.499 $\pm$ 0.039\\
    &  0.9& 0.45 &  10 &  0.514 $\pm$ 0.020 \\
    &       &         &  5 &  0.492 $\pm$ 0.022\\
    &       &         &  1 &  0.436 $\pm$ 0.032 \\
1.0 & 0.3 & 0.15 & 10 &  0.969 $\pm$ 0.139\\
    &       &         &  5 &  0.922 $\pm$ 0.132 \\
    &       &         &  1 &  0.998 $\pm$ 0.143\\
    &  0.6 & 0.3  &  10 &  1.007 $\pm$ 0.056\\
    &       &         &  5 &  0.960 $\pm$ 0.056\\
    &       &         &  1 &  0.988 $\pm$ 0.047 \\
    &  0.9& 0.45 &  10 &  0.952 $\pm$ 0.051\\
    &       &         &  5 & 0.973 $\pm$ 0.048 \\
    &       &         &  1 &  0.942 $\pm$ 0.061 \\

\hline
\end{tabular}
\label{tab:sim}
\vspace{10pt}
\end{table*}

In this study, we take advantage of the wealth of SNR data available in
the \chandra archive to investigate and quantify the
systematic errors present in the determination of the hydrogen column
density from the analysis of the X-ray data. We analyze all of the
observations from the \chandra archive using standardized procedures
so that we can quantify the systematic errors on each measurement.
Using only observations from the \chandra archive also ensures a
completely uniform data set: All observations were performed using the
ACIS detector, and each had spectra generated and analyzed using the
same treatment for background subtraction and model fitting routines
using the spectral analysis software {\it xspec} (version 12.8.1; \citealt{xspec}, with NEIVERS 1.1). This
consistent treatment of a uniform data set gives us the opportunity to
quantify the existing systematic errors in a way that had not been
possible before.

The uniformity of our data set also allows us to explore the
uncertainties associated with fitting spectra with a variety of
models. SNRs can be home to a wide range of plasma properties due to
the wide range of ages, host environments, and composition of the
sampled gas. In general, however, SNRs exhibit continuum emission
driven by thermal bremsstrahlung accompanied by emission lines produced by ejecta
material or swept up ISM. In the 0.5-5.0 keV range, where we
predominantly perform the spectral fits, these features manifest most
visibly in magnesium, silicon and sulphur lines (though features due
to iron, oxygen and neon are also possible within this
range). Non-thermal features can also be present due to particle acceleration
in SNR shocks. Due to the wide range
in spectral features and plasma conditions, there also exist a range
of models that can be used to fit X-ray spectra which will be detailed
in Section 2.

\begin{figure}
\centering
\includegraphics[scale=0.45]{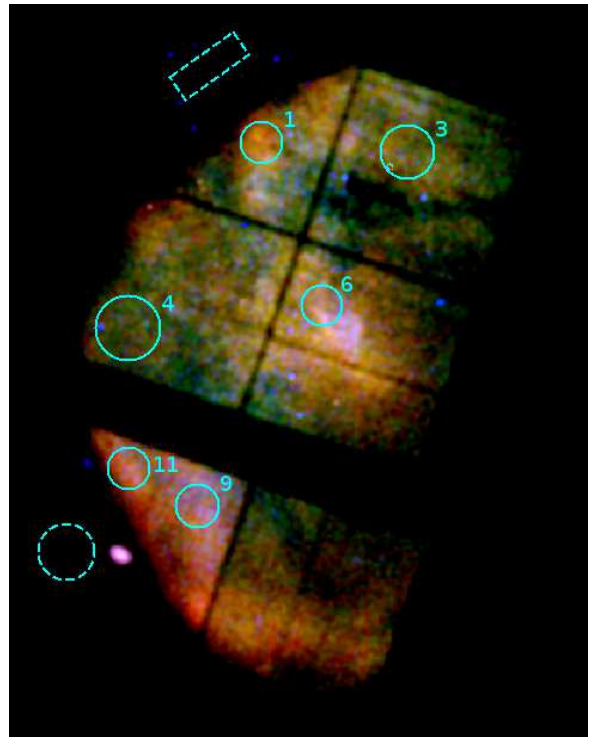}
\caption{A \chandra image of SNR G109.1-1.0 taken with the ACIS-I
  detector (data from OBSID 1901, \textit{red:} 0.5-1.1 keV;
    \textit{green:} 1.1-1.6 keV; \textit{blue:} 1.6-5.0 keV). The
  numbered circles show six of the 11 regions that were selected for
  spectrum extraction from this remnant. The dashed rectangle and
  circle shows the regions selected for background subtraction.}
\label{fig:snr_regions}
\end{figure}

Making use of the high spatial resolution and large number of counts
in the \chandra archival data, we also investigate the systematic
uncertainties associated with differing lines of sight towards larger
remnants. We assume that all absorption is due to interstellar gas and
dust along a line of sight, with no intrinsic absorption from the
optically thin remnant. This assumption allows there to be differing values for
$N_{\rm H}$ for different regions in a given SNR due to real
differences in gas along a line of sight, which is supported by deep
observations of large remnants such as W49B \citep{w49blegacy} and CTB
109 (also known as SNR G109.1-1.0; \citealt{ctblegacy}), for which
many regions can be fit. When the archival observations have
sufficient duration to have multiple high-count regions, we are able
to fit multiple regions to understand how the value of $N_{\rm H}$
varies over the remnant. By fitting multiple regions with a
  range of models, we can determine the magnitude of the total
  systematic error on $N_{\rm H}$ measurements in the direction of the
  SNR. We note that any dust that is intrinsic to the SNR contributes
negligible extinction.  The integrated column density of swept-up
dust is small, and that of dust condensed from the stellar ejecta
is smaller still. Moreover, SNR shocks are very efficient dust
destroyers \citep{temim2015}, making any contribution to the
overall extinction exceedingly small.

Finally, and possibly most importantly, many of the previous analyses
of \chandra and other SNR data used solar abundances of \citet{angr}
derived from observations of the sun and meteorites. \citet{wilms2000}
showed that the abundances in the ISM can differ from solar values,
which can affect the fit value of $N_{\rm H}$. It is also
  important to consistently use an ISM absorption model that features
  improved calculations of the absorption cross sections when seeking
  to improve systematic errors in the hydrogen column density
  determination.
We use the {\it wilms} abundance table and {\it tbabs} absorption
model within {\it xspec} in order to accurately model the interstellar
absorption to measure the hydrogen column density. The {\it tbabs} model in {\it xspec} adopts 
the cross-sections from \citet{verner1995}. This standard that
we adopt for the entire sample leads to larger ($\sim 30\%$) column
densities than those reported in the earlier studies.

In Section~\ref{sec:motivation}, we discuss the variety of models used
to fit our data, as well as the results of simulated data sets.
Sections~\ref{sec:methods} and~\ref{sec:results} detail the processes we
used to process the data, as well as our approach to identifying
systematic errors and determining total uncertainties. Finally, in
Section~\ref{sec:application} we use our new data set to derive the
slope of the $N_{\rm H}$-$A_{\rm V}$ relation.

\begin{figure}
\centering
\includegraphics[scale=0.45]{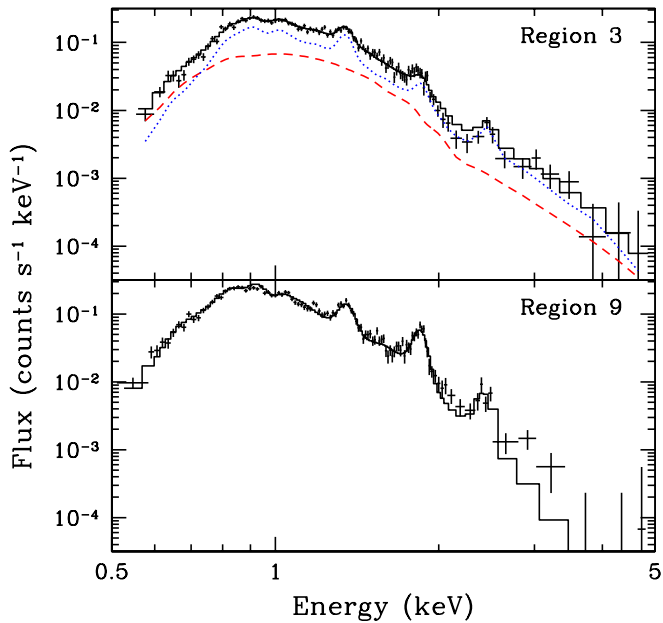}
\caption{Thermal spectra extracted from two different regions of SNR
  G109.1-1.0 (corresponding to regions 3 and 9 shown in
  Figure~\ref{fig:snr_regions}) in the $0.5-5.0$~keV range
  with best fit model components overlaid. Region 3 spectrum
  exhibits weak emission lines (from the thermal component
    shown in the dotted blue line) with some power-law component
  contribution (shown in the red dashed line), while region 9
  spectrum has stronger lines and no power-law component
  (thermal model shown as solid black line).}
\label{fig:spectra}
\vspace{5pt}
\end{figure}

\section{Models and Simulated Results}
\label{sec:motivation}

The models we used to fit each thermal region ($raymond$, $mekal$,
$nei$, and $pshock$) were chosen to span the range of complexity one
might expect from a region of plasma in an SNR. The $raymond$ model
\citep{raymond} is the simplest, modeling a hot, diffuse gas in
collisional equilibrium. Similarly, the $mekal$ model also calculates
spectra of a plasma in collisional equilibrium, but with improved
handling of the Fe-L complex \citep{mekal}. The $nei$ and $pshock$
models \citep{nei} are more complicated, providing spectra for a
non-equilibrium ionization (NEI) collisional plasma ($nei$), and constant
temperature NEI plasma heated by a plane-parallel shock ($pshock$). For
non-thermal regions, the models $powerlaw$ and/or $srcut$ were
used. Both models produce a spectrum we would expect from synchrotron
emission from a power-law distribution of shocked electrons interacting
with an SNR's magnetic field, with an additional exponential cutoff of
the distribution for the $srcut$ model \citep{powerlaw,srcut}.

As was discussed previously, differing lines of sight can sample
different paths through the ISM, so there is a chance for real
variations in $N_{\rm H}$ measurements across an SNR. There can also be
small systematic variations introduced by fitting the regions with the
models discussed above. Each region may be composed of multiple plasma
components, so fitting to a single plasma model may introduce a
systematic error or a bias. To show this, we performed fits to
simulated spectra produced with a range of $N_{\rm H}$. We produced
the simulated spectra using the $sedov$ model in {\it xspec}
(version 12.8.1 using NEIVERS 1.1), which models the
  total emission from an SNR undergoing adiabatic expansion
\citep{nei}.  The complete set of simulated spectra cover a grid of
inputs for shock temperature
($T_s=0.3-0.9$~keV), electron temperature
($T_{es}=0.15-0.45$~keV), ionization age
($\tau_0 = 1-10 \times 10^{11}$ s
  cm$^{-3}$), and $N_{\rm H}$
($0.5-1.0 \times 10^{22}$
  cm$^{-2}$). For each parameter set, we simulated 200
spectra and then binned these to have at least 50 counts per bin. We
fit each spectrum with a $pshock$ model, using the \textit{tbabs}
absorption model with the $wilms$ abundance table. We present the
average of the best fit values for each parameter set in
Table~\ref{tab:sim}, along with the standard deviation of all fit
values.

The results in Table~\ref{tab:sim} show that while a single plasma
model approximately recovers the assumed $N_{\rm H}$ value from a
complex plasma, it can introduce significant statistical
  errors and suggests a possible bias. In particular, the $N_{\rm H}$
values we measure from the simulated data are close to but
often lower than the assumed values. This is in fact not
surprising: fitting a single-temperature model to a simulation that is
meant to represent the entire remnant (i.e., a Sedov spectrum
which integrates over a range of temperatures) cannot
adequately describe the spectrum or yield a very accurate value of the
hydrogen column density.
 
To minimize these effects in the analysis of the actual data,
we take a three-pronged approach. First, we extract spectra from as
small regions of the remnant as possible, to avoid creating
complicated, multi-temperature regions with multiple plasma
components.  Second, we perform fits with thermal, non-thermal, and
mixed thermal/non-thermal models, to capture the spectral
characteristics of the regions correctly. Third, we use Bayesian
statistical tools (discussed in Section~\ref{sec:bayes}) to combine
measurements from different spatial regions as well as from different
spectral model results to assess any systematic uncertainties in the
$N_{\rm H}$ measurement for a given remnant. It is important to 
note, however, that the simulated spectra results on their own are not the 
sole motivation for the Bayesian analysis; if we believed that the simulated 
results were immediately comparable to the \chandra data results, then we 
could use the simulated results to establish the magnitude of a bias term. 
In Section~5, we will use the total uncertainties we determine for the $N_{\rm H}$ towards each
remnant to more accurately constrain the $N_{\rm H}/A_{\rm V}$
relationship.

\section{Chandra Data Pipeline and Processing}
\label{sec:methods}

\subsection{Pre-Processing}

For each of the selected SNRs, we downloaded archival data from the
Chandra archive. For remnants with multiple pointings and exposures,
we chose one observation to eliminate any need to co-add multiple
exposures, and placed preference on newer observations that were long
enough to provide multiple regions with sufficient counts for spectral
analysis. We ran the data through two pipeline scripts that were
designed to automate the processing procedures. All scripts utilized
\textit{CIAO} tools used version $4.5$. Our first script reapplied the
latest calibration and produced a new level=1 event file. Because the
majority of remnants were taken with exposure mode FAINT or VFAINT, we
cleaned the ACIS background using procedures that were appropriate for
these modes.  Finally, we filtered the level=1 file for bad grades and
applied the good time intervals to generate the new level=2 event
file.

\begin{table*}
\centering
\caption{Example Fit Values for SNR G109.1$-$0.0}
\begin{tabular}{l l l c c}
\hline
\hline
Region & Model & Thawed Elements & Fit $N_{\rm H}$ & $\chi^{2}/\nu$ \\
 & & & $(10^{22}\ cm^{2})$ & \\
\hline
\hline
1 (NE) & vnei + powerlaw & Mg, Si & $1.07^{+0.07}_{-0.07}$ & 146.94/97 \\
 & vpshock + powerlaw & Mg, Si, S & $1.05^{+0.08}_{-0.07}$ & 143.81/96 \\
3 (N) & nei + powerlaw & None & $0.95^{+0.09}_{-0.09}$ & 132.71/111 \\
 & vnei + powerlaw & Mg, Si & $0.93^{+0.10}_{-0.16}$ & 130.55/109\\
 & pshock + powerlaw & None & $0.95^{+0.09}_{-0.04}$ & 131.83/111\\
 & vpshock + powerlaw & Mg, Si & $0.89^{+0.09}_{-0.06}$ & 126.88/109 \\
4 (E) & vnei + powerlaw & Mg, Si & $0.90^{+0.07}_{-0.07}$ & 160.49/124 \\
 & vpshock + powerlaw & Mg, Si & $0.93^{+0.08}_{-0.07}$ & 153.14/124 \\
6 (Center) & vnei & Mg, Si, S & $0.86^{+0.08}_{-0.07}$ & 138.73/110 \\ 
 & vpshock & Mg, Si, S & $0.71^{+0.08}_{-0.05}$ & 130.71/110 \\
9 (SE Inner) & vnei & Mg, Si, S & $1.24^{+0.06}_{-0.07}$ & 139.21/117 \\
 & vpshock & Mg, Si, S & $1.23^{+0.05}_{-0.05}$ & 137.94/117 \\
11 (SE Outer) & nei & None & $0.87^{+0.06}_{-0.06}$ & 142.52/111 \\
 & vnei & Mg, Si & $0.88^{+0.08}_{-0.07}$ & 142.15/109 \\
 & pshock & None & $0.88^{+0.08}_{-0.06}$ & 130.27/111 \\
 & vpshock & Mg, Si & $0.89^{+0.08}_{-0.08}$ & 129.88/109 \\
\hline
\end{tabular}
\label{tab:fullresults}
\vspace{10pt}
\end{table*}

\subsection{Region Selection and Data Processing}

We used the new level=2 event file to select regions for spectral
extraction. We selected regions using ds9 and chose areas that had
$\ge 10,000$ counts, as well as sampling a variety of lines-of-sight
across the remnant. In some remnants, it was preferable or necessary
to select regions with non-thermal emission. We processed these using
procedures identical to the thermal regions, but fit them with
non-thermal models. In general, regions were selected using
  the morphology of the remnant as a guide, such that the regions
  would contain one variety of plasma (the importance of which is
  highlighted in Section 2). We extracted a spectrum from each region
and then grouped it to a minimum of 25 counts per bin.

As a typical example of the regions selected for a remnant, we show
SNR~G109.1$-$1.0 in Figure~\ref{fig:snr_regions}, with a number of the
selected regions marked with circles and the regions selected for
 local background shown with a dashed circle and rectangle. In the
full analysis of G109.1$-$1.0, we used a total of eleven regions.
However, for clarity and simplicity, we chose to highlight only six in
this figure. We show the resultant binned spectra from two of these
regions in Figure~\ref{fig:spectra}, again as two representative
examples of the types of spectra we encountered in the analyses.
Note that small contributions from model components (such as the 
power-law component in Region 3 of Figure~\ref{fig:spectra}) may appear
insignificant, but are important for producing acceptable fits. 

We fit a variety of models (discussed in Section~\ref{sec:motivation})
to each grouped spectrum. We fit each model with default ISM
  abundances, but it was often necessary to allow some of the plasma
  abundances to vary in order to properly fit regions with
  ejecta-enriched plasma. The number and variety of the free
  abundances were allowed to change from model to model for a given
  region, and only the plasma models' abundances were allowed to
  vary. For non-thermal regions, this process was the same, but
simplified by the fact that there were no free abundance parameters
possible for those models.

Despite our best efforts to select regions that would be composed of
only one variety of plasma (and thus could be fit using one model),
sometimes this was infeasible (for example, if the regions had to be
large in order to contain an adequate number of counts). In such cases
we needed to add other components to the models. Most commonly, this
required adding a nonthermal model to the thermal model to compensate
for regions near nonthermal filaments or a central pulsar wind nebula
(PWN).  However, in some cases we added Gaussian features to fit
spectral lines that were either not properly fit or non-existent in
the thermal model, or to account for ejecta mixing into a non-thermal
region, causing emission lines to appear on top of a power-law
spectrum. For the most part, the data were good enough that we could
find regions that allowed for a single model to produce an acceptable
fit, but when included, these additional components did not make
statistically significant changes in the best-fit value of $N_{\rm
  H}$.

\begin{figure*}
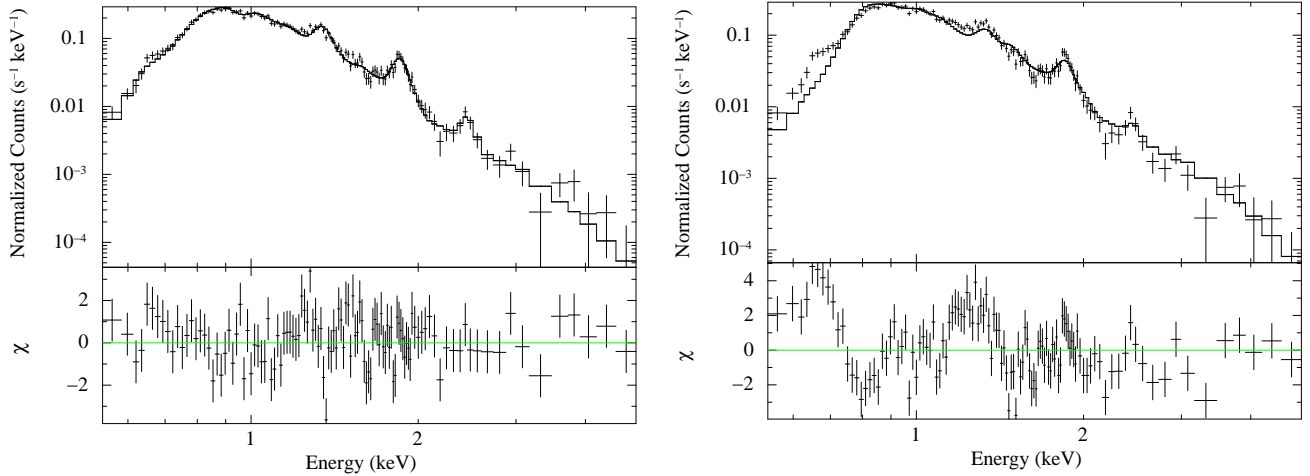

\centering
\centerline{%
\includegraphics[scale=0.35,angle=270]{f3a.eps}%
\includegraphics[scale=0.35,angle=270]{f3b.eps}%
}%
\caption{Two example model fits to a spectrum of SNR G109.1-1.0
  extracted from Region 6. The acceptable fit (left panel) is a $vnei$
  model, whereas the unacceptable fit (right panel) is a $vraymond$
  model in this particular example. Each model has Mg, Si, and S
  abundances as free fit parameters.}
\label{fig:compare}
\end{figure*}

We determined if a spectral fit was good both by visually checking the
final fit, as well as by using the $\chi^{2}/\nu$ fit
statistic.  We show in Figure~\ref{fig:compare} an example of an
acceptable fit compared to a poor fit. We did not place a hard upper
limit on the fit statistic because some large
$\chi^{2}/\nu$ values were dominated by a single feature
that was not captured by the model but did not affect the inferred
$N_{\rm H}$. Nevertheless, nearly all of the fits used in our final
analysis had $\chi^{2}/\nu \le 2$.

\section{Results}
\label{sec:results}

\subsection{Results of The Spectral Analysis}

For each SNR region, we included all models that produced acceptable
fits. As a typical example, we show in Table~\ref{tab:fullresults} the
complete fit results for six of the regions of G109.1$-$1.0. It is
evident from this table that $vnei$ and $vpshock$ models with thawed
abundances were commonly the best models for remnants with thermal
regions, with a $powerlaw$ component added in when necessary. The
$raymond$ and $mekal$ models ranged from fairly successful to very
poor, which presumably reflects significant non-equilibrium ionization
contributions in the data. For SNRs with completely non-thermal
regions, the $powerlaw$ model was the default model with consistent
success (Two of the remnants, G120.1+1.4 and G04.5+6.8, were better
fit by an $srcut$ model). All errors presented on the fit $N_{\rm H}$
values in Table~\ref{tab:fullresults} are 90\% confidence ranges. The
complete fit results for all remnants used in our analysis have been
released as a Zenodo dataset (DOI: 10.5281/zenodo.17183).

\begin{table}[htb]
\centering
\caption{Measured Column Densities for All SNRs}
\begin{tabular}{l r l l}
\hline
\hline
SNR & Number of & $N_{\rm H}$ & Error \\
 & Regions/Models & \multicolumn{2}{c}{$(10^{22}~\rm{cm}^{2})$} \\
\hline
\hline
G0.0+0.0 & 3 & 9.86 & 0.34 \\
G04.5+6.8 & 4 & 0.54 & 0.011 \\
G06.4-0.1 & 22 & 0.63 & 0.19 \\
G53.6-2.2 & 2 & 0.67 & 0.03 \\
G54.1+0.3 & 1 & 2.55 & 0.04 \\
G69.0+2.7 & 1 & 0.45 & 0.02 \\
G109.1-1.0 & 30 & 0.90 & 0.14 \\
G111.7-2.1 & 8 & 1.17 & 0.16 \\
G116.9+0.2 & 2 & 0.92 & 0.07 \\
G119.5+10.2 & 1 & 0.38 & 0.11 \\
G120.1+1.4 & 14 & 0.80 & 0.10 \\
G130.7+3.1 & 6 & 0.54 & 0.01 \\
G184.6-5.8 & 1 & 0.30 & 0.02 \\
G260.4-3.4 & 3 & 0.35 & 0.15 \\
G263.2-3.3 & 1 & 0.03 & 0.01 \\
G327.6+14.6 & 2 & 0.16 & 0.01 \\
G332.4-0.4 & 9 & 0.87 & 0.35 \\ 
\hline
\end{tabular}
\label{tab:summary}
\end{table}

\subsection{Bayesian Analysis to Determine $N_{\rm H}$ and Its Uncertainty}
\label{sec:bayes}

Each of the fits within a region have formal (and often asymmetric)
uncertainties, as we showed in Table~\ref{tab:fullresults} for
SNR~G109.1$-$1.0. From these fits, the dispersion in the
measurements arising from spatial sampling or from different continuum
models can be calculated. For a number of remnants, this contribution is comparable to the formal uncertainties. In others, however,
the difference between the various measurements of $N_{\rm H}$ is
significantly larger than the formal uncertainties, pointing to
systematic uncertainties originating from spatial sampling, from the
choice of model, or both. For the purposes of our analysis, we
  do not distinguish between the variance of the ISM along different
  lines of sight and the systematic error introduced by the incomplete
  plasma models and we combine these different types of systematic
  error into a single term (see, e.g., \citealt{sinervo2003}). Our aim
  in this section is to determine the most likely $N_{\rm H}$ value
  for each remnant, as well as a measure of the combined formal and
  systematic uncertainties in this quantity. We accomplish this by
  finding the parameters of the underlying $N_{\rm H}$ distribution
  for each remnant that is consistent with our sample of
  measurements. We use this distribution to find the most likely value
  of the hydrogen column density and its uncertainty.

We start with a parametric form of the $N_{\rm H}$ distribution and
use the set of measurements from the individually fitted
  regions in order to estimate its parameters. We take the assumed
underlying distribution to be a Gaussian
\begin{equation}
P(N_{\rm H}; \sigma, N_{\rm Hc}) = C \exp\left[-\frac{(N_{\rm H}-N_{\rm Hc})^2}
{2\sigma^2} \right]
\label{eq:NH_distr}
\end{equation}
with a mean $N_{\rm Hc}$ and a standard deviation $\sigma$ that
can be different for each remnant. In this and the following
expressions, $C$ is a proper normalization constant such that
\begin{equation}
\int_0^\infty P(N_{\rm H}; \sigma, N_{\rm Hc}) dN_{\rm H} = 1.
\end{equation} 

We also need to model the individual measurement uncertainties
$P_i(N_{\rm H})$, where $i$ represents a particular SNR region/model
combination that yields a single measurement of $N_{\rm H}$
for that region.  In general, the $\chi^2$ surface for each
measurement is not simple or symmetric around the minimum, leading to
asymmetric formal errors in these individual measurements. However,
around the minimum, it is accurate to represent the likelihood using
two half Gaussians with different dispersions that smoothly connect at
the most likely column density $N_{H0,i}$ for each measurement; i.e.,
\begin{align} & P_i({\rm data} \vert N_{\rm H}) = \nonumber \\ & \left\{
\begin{array}{ll} C_i \exp\left[-\frac{(N_{\rm H}-N_{\rm H0,i})^2}
{2\sigma_{-,{\rm NH,i}}^2} \right], & \;\; N_{\rm H} < N_{\rm H0,i} \\
C_i \exp\left[-\frac{(N_{\rm H}-N_{\rm H0,i})^2} {2\sigma_{+,{\rm
NH,i}}^2} \right], & \;\; N_{\rm H} > N_{\rm H0,i} \end{array}\right.
\label{eq:prob_data} 
\end{align}

\noindent where ``data'' stands for the most likely column density $N_{H0,i}$  and the 
two associated uncertainties $\sigma_{-,N_H,i}$ and $\sigma_{+,N_H,i}$
for the $i$-th SNR region/model combination.

We want to calculate the quantity $P(\sigma,N_{\rm Hc}\vert {\rm
  data})$, which measures the posterior likelihood of the
parameters of the $N_{\rm H}$ distribution, given the
observations. Using Bayes' theorem, we can write this as
\begin{equation}
P(\sigma,N_{\rm Hc}\vert {\rm data}) = C_2 P({\rm data}\vert \sigma, N_{\rm Hc})
P(\sigma)P(N_{\rm Hc})\;,
\label{eq:bayes}
\end{equation}
\noindent where $C_2$ is a normalization constant and $P(\sigma)$ and
$P(N_{\rm Hc})$ are the priors over the values of the Gaussian
dispersion, $\sigma$, and the peak of the $N_{\rm H}$ distribution,
$N_{\rm Hc}$. Here, ``data'' stands for the ensemble of the $N_{H0,i},\  
\sigma_{-,N_H,i}$, and $\sigma_{+,N_H,i}$ values for a 
particular SNR. We take a flat prior over the Gaussian dispersion
$\sigma$ between $\sigma_{\rm min}$ that is equal to 0.1 times the
smallest formal uncertainty obtained from a spectral fit for each
remnant and $\sigma_{\rm max}$ that is equal to 10 times the largest
difference between two $N_{\rm H}$ measurements for each remnant:
\begin{equation}
P(\sigma) = \left\{\begin{array}{ll}
0, & \sigma \le \sigma_{\rm min}\\
\frac{1}{\sigma_{\rm max}-\sigma_{\rm min}}, & \sigma_{\rm min}<\sigma<\sigma_{\rm max}\\
0, & \sigma>\sigma_{\rm max}\;.
\end{array}\right.
\label{eq:priors}
\end{equation}
Similarly, we take a flat prior over the centroid of the $N_{\rm H}$
distribution $N_{\rm Hc}$ that spans the range from 0.1 times the
smallest $N_{\rm H}$ measurement to 10 times the largest $N_{\rm H}$
measurement per source. These limits ensure that the particular
minimum and maximum values of the prior distribution do not affect the
results.

In equation~(\ref{eq:bayes}), the quantity $P({\rm data}\vert \sigma,
N_{\rm Hc})$ measures the likelihood that we will make a particular
set of measurements for the column density given the values of the
parameters of the column density distribution. We need to estimate
this quantity, given the measurement likelihoods given in
equation~(\ref{eq:prob_data}). We will assume that each measurement is
independent, so that
\begin{align}
& P({\rm data}\vert \sigma, N_{\rm Hc}) = \nonumber \\
& \quad \prod_i \int dN_{\rm H} P_i({\rm data}| N_{H}) P(N_{\rm H};\sigma, N_{\rm Hc})\;.
\label{eq:product}
\end{align}
Combining this last equation with equation~(\ref{eq:bayes}), we obtain
the posterior likelihood
\begin{align}
& P(\sigma,N_{\rm Hc}\vert {\rm data}) =  C P(\sigma)P(N_{\rm Hc}) \times \nonumber \\
& \quad \prod_i \int dN_{\rm H} P_i ({\rm data}| N_{\rm H}) P(N_{\rm H};\sigma, N_{\rm Hc})\;, 
\label{eq:final}
\end{align}
where $C$ is the overall normalization constant. We can use equation~\ref{eq:final} to determine the parameters $\sigma$ and $N_{\rm Hc}$ given the individual region/model fits for each remnant. The dispersion ($\sigma$) is then a measure of the systematic uncertainty associated with the remnant. If a remnant has a significant systematic uncertainty, that contribution is considered when we define the most likely value of the $N_{\rm H}$ for the SNR and its uncertainty.   

\begin{figure*}
\centering
\centerline{%
\includegraphics[scale=0.42]{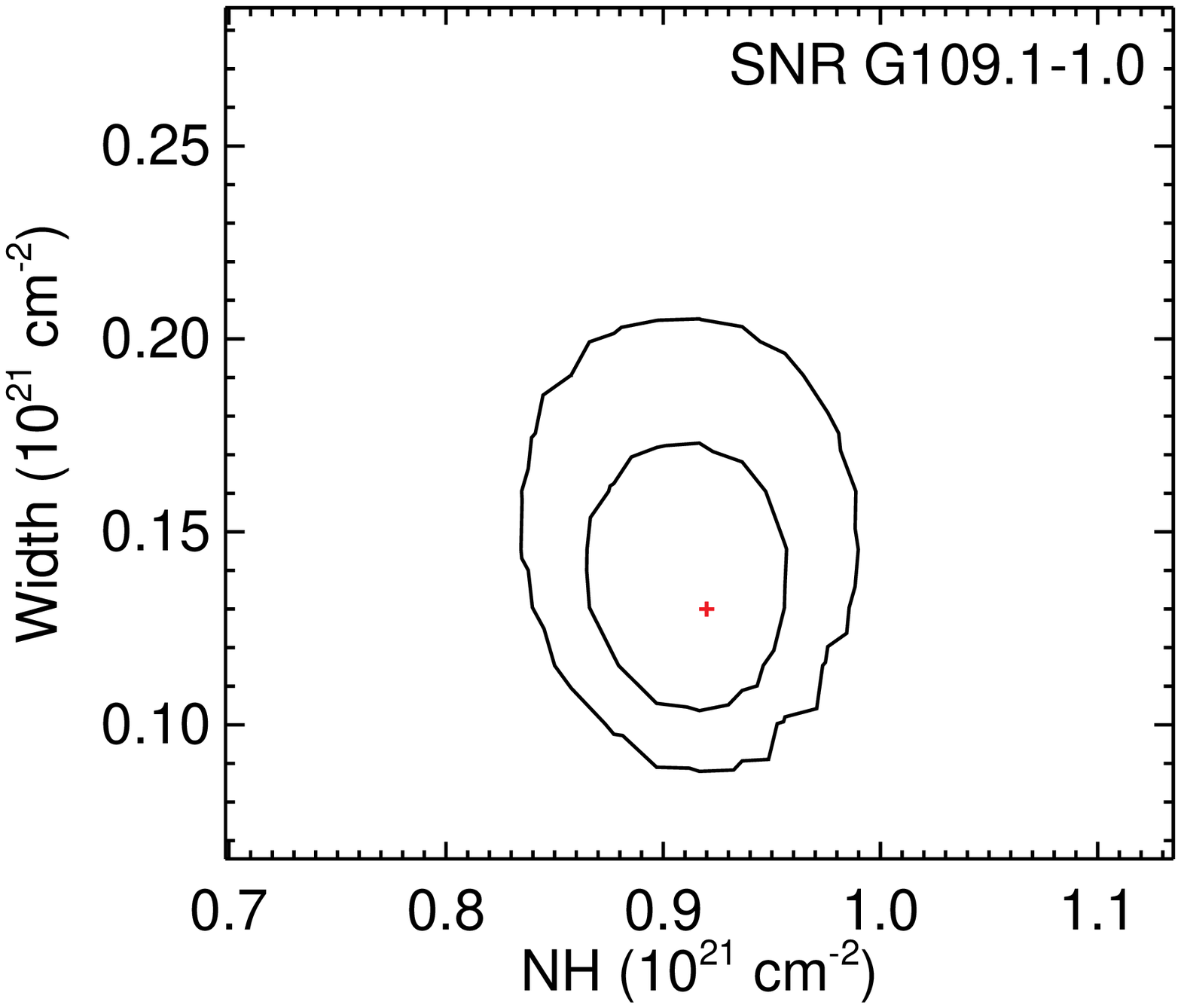}%
\includegraphics[scale=0.42]{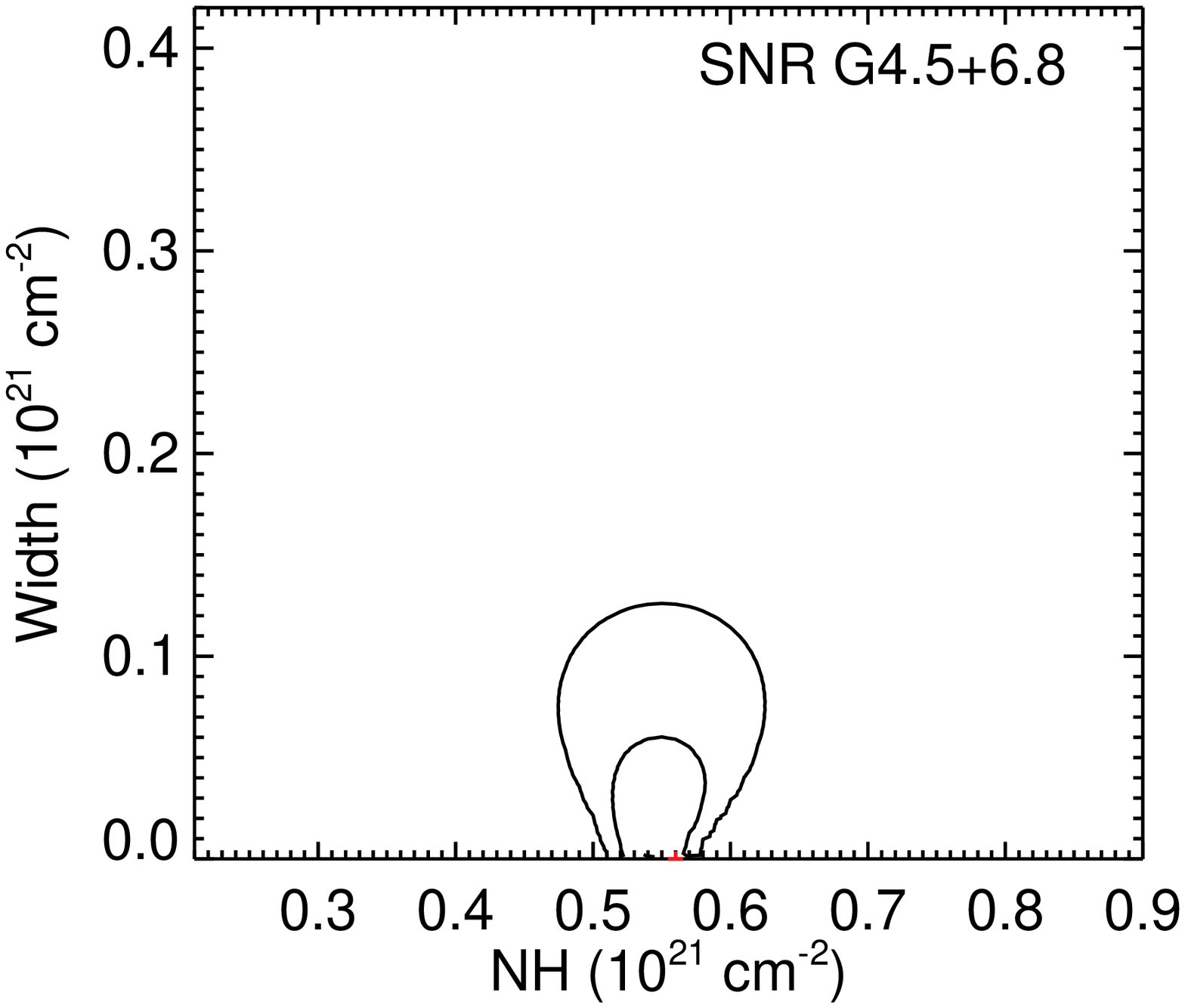}%
}%
\caption{$p=0.32$ and $p=0.05$ Bayesian credible regions for the central
  value and the dispersion of an underlying Gaussian distribution of
  $N_{\rm H}$ measurements for SNRs G109.1-1.0 (left) and G04.5+6.8
  (right).  The dispersion of the distribution reflects the systematic
  uncertainty in the measurements.  The dispersion contours for
  G04.5+6.8 are consistent with 0.0, which indicates that systematic
  uncertainties are not significant. For G109.1$-$1.0, the most likely
  dispersion is not consistent with zero, indicating a level of
  systematic error comparable to the formal errors. These are included
  in the uncertainties in Table~\ref{tab:summary} following the
  discussion in Section~4.}
\label{fig:conf_contours}
\end{figure*}

In Figure~\ref{fig:conf_contours}, we show as examples the
  Bayesian credible regions over the parameters of the $N_{\rm H}$
distribution that correspond to the measurements for G109.1$-$1.0 and
G4.5+6.8.  We show with a dot the most likely values of the centroid
and dispersion; i.e, the peak of the posterior likelihood given by
equation~(\ref{eq:final}). In the case of G4.5+6.8, the best-fit value
of the dispersion is very close to zero and this parameter is
consistent with being zero within the $p=0.05$ credible region, indicating
a negligible level of systematic uncertainty. On the other hand, for
remnants such as G109.1$-$1.0, the best fit value of the dispersion is
not consistent with being zero, indicating the presence of systematic
uncertainties arising from spatial sampling across the remnant or the
differences among spectral models.

The previous Bayesian analysis to quantify the systematic error term yields two categories of remnants. 
In the first category the systematic dispersion ($\sigma$) is
consistent with zero, so we assign a zero systematic uncertainty, i.e.,
set $P(N_{\rm H}; \sigma, N_{\rm Hc}) = \delta(N_{\rm H}-N_{\rm Hc})$
in equation~(\ref{eq:product}). This allows us to find a properly
weighted average of the $N_{\rm H}$ measurements using their formal
uncertainties and report the best-fit value and its formal uncertainty
in Table~\ref{tab:summary}. In the second category of remnants, we
compute the posterior likelihood over their $N_{\rm H}$ by weighing
each Gaussian distribution with a given centroid $N_{Hc}$ and
dispersion $\sigma$ with the likelihood calculated in
equation~(\ref{eq:final}) that those pair of parameters represent the
observed $N_{\rm H}$ values:
\begin{align}
& P(N_{\rm H}) = \nonumber \\
& \quad \int \int P(N_{\rm H}; \sigma, N_{\rm Hc}) P(\sigma,N_{\rm Hc}\vert {\rm data}) 
dN_{\rm Hc} d\sigma
\end{align}
We use this distribution to infer the best-fit values of $N_{\rm H}$
and its uncertainty for each source and report this in
Table~\ref{tab:summary}.

\break

\section{Application to the $N_{\rm H}$/$A_{\rm V}$ Relation}
\label{sec:application}

We presented in the previous section the hydrogen column densities
toward a large sample of SNRs that we measured using the \chandra data
archive. We now combine these $N_{\rm H}$ measurements with the
optical extinction measurements presented in \citet{go2009} for our
sample of SNRs to determine a relation between these two quantities
that accounts for systematic errors. The optical extinction toward the
remnants in our sample was determined through several different
methods, which we list in Table~\ref{tab:av}. Most of these methods
involve measuring the intensity ratio of two emission lines, for which
the intrinsic ratio is known. Comparing the observed intensity ratio
to the intrinsic ratio yields a reddening, which is then converted
into a measurement of the optical extinction. These methods are
challenging, in general, because a high signal-to-noise ratio spectrum
is necessary to obtain the intensity ratios.

Several emission line pairs are frequently used for this purpose.
$H_\alpha$ (6563 \AA) to $H_\beta$ (4861 \AA) line ratio, referred to
as the Balmer decrement method, is one of the most well known and
reliable ones among these pairs.  Other emission line ratios from the
SII multiplet \citep{miller1968} and FE[II] IR transitions
\citep{oliva1989} are also used in a similar way.  For all of these
pairs of lines, the ratio depends only very weakly on the temperature
and density of the emitting plasma, leading to minimal uncertainties
in the calculation of the ratio \citep{osterbrock1989,lequeux2005}.

\begin{table}[b]
\centering
\caption{$A_{\rm V}$ Values from \citet{go2009}}
\begin{tabular}{l l l l c}
\hline
\hline
SNR & $A_{\rm V}$ & Error{$^{a}$} & Method & Ref. \\
 & (\textit{mag}) & (\textit{mag}) &   &  \\
\hline
\hline
G0.0+0.0 & 29 & 2 & Nearby Stars & (1)\\
G04.5+6.8 & 2.5 & 0.9 & FeII Ratio & (2) \\
G06.4-0.1 & 3.57 & 0.47 & H$_{\alpha}$/H$_{\beta}$ & (3) \\
G53.6-2.2 & 3.57 & 0.47 & H$_{\alpha}$/H$_{\beta}$ & (3) \\
G54.1+0.3 & 8.0 & 0.70 & Nearby Stars & (4) \\
G69.0+2.7 & 2.48 & -- & H$_{\alpha}$/H$_{\beta}$ & (5) \\
G109.1-1.0 & 3.15 & 0.65 & H$_{\alpha}$/H$_{\beta}$ & (6) \\
G111.7-2.1 & 5.0 & 0.40 & SII ratio & (7)\\
G116.9+0.2 & 2.70 & 0.5 & H$_{\alpha}$/H$_{\beta}$ & (8) \\
G119.5+10.2 & 1.27 & 0.41 & Extinction Map & (9) \\
G120.1+1.4 & 1.86 & 0.12 & Nearby Stars & (10) \\
G130.7+3.1 & 2.11 & -- & Nearby Stars & (11) \\
G184.6-5.8 & 1.55 & 0.186 & Ly$\alpha$ Absorption & (12)\\
G260.4-3.4 & 2.60 & -- & Nearby Stars & (13)\\
G263.2-3.3 & 0.38 & -- & H$_{\alpha}$/H$_{\beta}$ & (14) \\
G327.6+14.6 & 0.34 & -- & HI/GC & (15) \\
G332.4-0.4 & 4.70 & 0.90 & FeII Ratio & (2) \\
\hline
\end{tabular} \\

\begin{flushleft}
{$^{a}$}Errors designated as '--' are taken as 15\%.

References: (1) \cite{predehl1994}; (2) \cite{oliva1989}; (3)
\cite{long1991}; (4) \cite{koo2008}; (5) \cite{hester1989}; (6)
\cite{fesen1995}; (7) \cite{hurford1996}; (8) \cite{fesen1997}; (9)
\cite{mavro2000}; (10) \cite{ruiz2004}; (11) \cite{fesen1988}; (12)
\cite{sollerman2000}; (13) \cite{gorenstein1975}; (14)
\cite{manchester1978}; (15) \cite{raymond1995}.
\end{flushleft}
\label{tab:av}
\end{table}

\begin{figure*}
\centering
\includegraphics[scale=0.75]{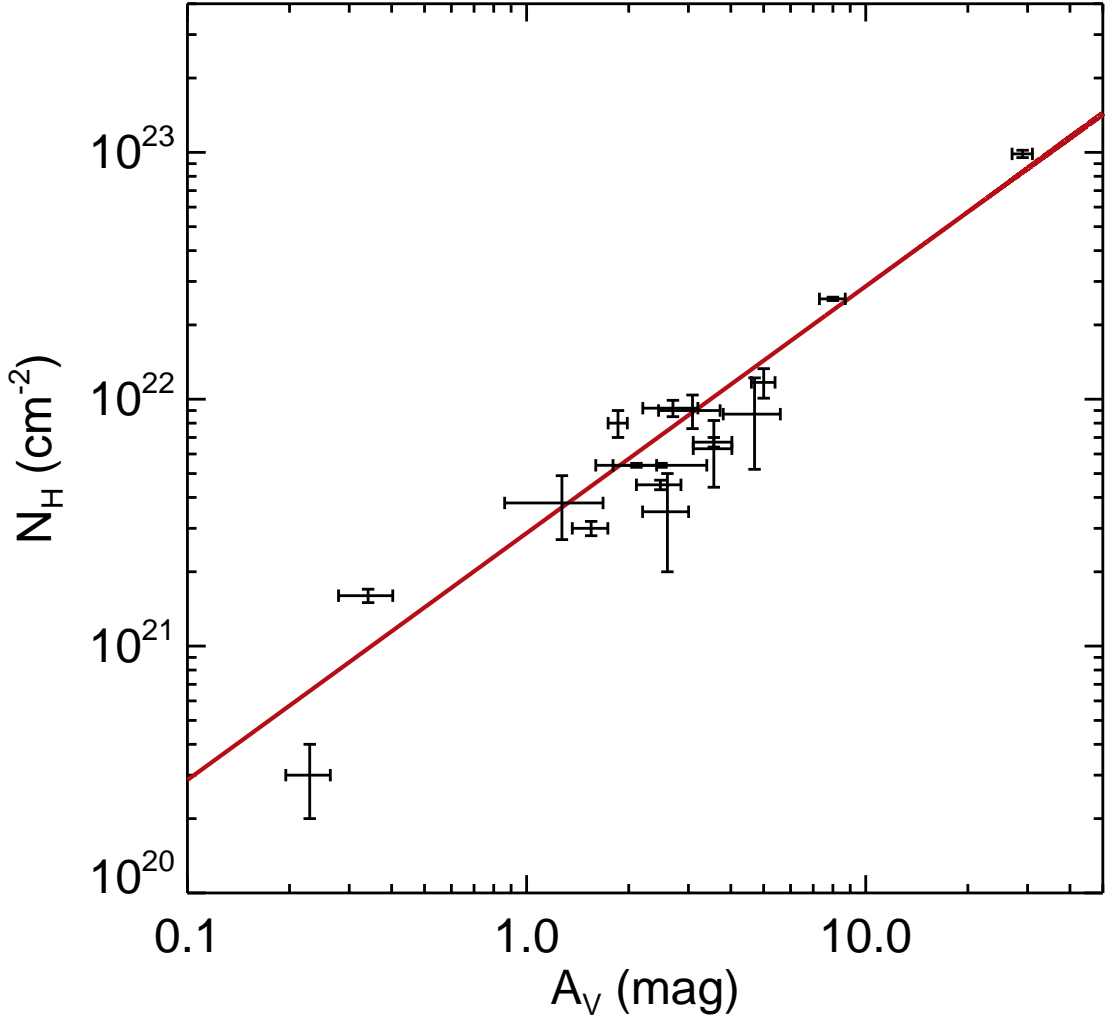}
\caption{The $N_{\rm H}$ and $A_{\rm V}$ measurements for the SNRs in
  the {\it Chandra} archive along with the best fit linear model,
  which gives $N_{\rm H} = 2.87\pm0.12\times10^{21}\ A_{\rm V}$.  The
  $N_{\rm H}$ measurements are presented in Table~\ref{tab:summary}
  and have a minimum error of 5\%, while the $A_{\rm V}$ measurements
  (listed in Table~\ref{tab:av}) are taken from \citet{go2009}. }
\label{fig:fit}
\end{figure*}

Another common method is to use stars near an SNR to estimate the
extinction toward that remnant. This method relies on having an
existing estimate of the distance to the remnant, and then identifying
stars at similar distances.  If suitable stars can be observed, the
extinction measurements of those stars can be applied to the remnant
\citep{koo2008, ruiz2004}.

In Figure~\ref{fig:fit}, we plot the hydrogen column density
measurements and their uncertainties presented in
Table~\ref{tab:summary} against the measurements of the optical
extinction summarized in Table~\ref{tab:av}. Following \citet{go2009},
we assign a 15\% error to the optical extinction measurements for
which uncertainties have not been reported (denoted by '-' in
Table~\ref{tab:av}). In addition, upon inspecting the uncertainties of
individual measurements in Table~\ref{tab:summary}, we note that a
best-fit line between these two quantities will be heavily influenced
by a small number of remnants where $N_{\rm H}$ measurements have very
small ($< 5\%$) formal uncertainties. Even though the sample of
regions and the range of models we considered in the spectral fits did
not allow us to measure a systematic uncertainty for these remnants,
considering the many possible sources of systematic uncertainty that
are usually present in the determination of the hydrogen column
densities, we assign a 5\% error to the $N_{\rm H}$ measurement of
SNR~$G130.7+3.1$, SNR~$G69.0+2.7$, SNR~$G004.5+06.8$, SNR~$G53.6-2.2$,
SNR~$G54.1+0.3$, and SNRG~$00.0+0.0$ to avoid biasing the results.

Using these data, we obtain a best-fit linear relation between the hydrogen
column density and the optical extinction that is described by 
\begin{equation}
N_{\rm H} = 2.87\pm0.12\times10^{21}\ A_{\rm V}~{\rm cm}^{-2}
\end{equation}
\noindent where $A_{\rm V}$ is in magnitudes and the error represents
the 1-$\sigma$ uncertainty. We show the best-fit line in
Figure~\ref{fig:fit}.

We can test the effect of imposing a floor to the $N_{\rm H}$
uncertainties by also fitting to the raw values. This increases the
best fit value of the $N_{\rm H}$/$A_{\rm V}$ to
$~(2.92\pm0.11)\times10^{21}$~cm$^{-2}$, which is within 1-$\sigma$ of
the best fit value presented above, but dominated by remnants with
small errors.  Finally, we note that the remnants with either very low
or high $N_{\rm H}$ will have the greatest leverage on the final fit.
To test the magnitude of this effect, we remove the points with the
highest and lowest values of the hydrogen column density. This results
in a best fit value of $~(2.76\pm0.13)\times10^{21}$~cm$^{-2}$, again
within 1-$\sigma$ of the best fit value using all of the data and a
minimum error imposed on the $N_{\rm H}$ measurements.

These results indicate that a consistently analyzed sample of SNRs
using ISM abundances for spectral fitting produces an $N_{\rm
  H}$/$A_{\rm V}$ relation that is significantly higher than previous
estimates. This has the effect of increasing the $N_{\rm H}$ derived
from existing E(B-V) measurements. Conversely, using the new relation
to estimate the brightness of an optical counterpart of an X-ray
source will result in less extinction compared to estimates calculated
with previous values of the $N_{\rm H}$/$A_{\rm V}$ relation. As
anticipated by \citet{go2009} and further discussed by
\citet{watson2011}, the primary reason for this difference is the
change in the abundances that are used for the ISM. The default
metallicity library of the frequently used spectral analysis package
{\it xspec} uses the solar abundances given by \citet{angr} and this
is indeed the main library that is utilized in the studies of the
spectral properties of X-ray sources. However, these values are known
to be on average 45\% higher than the updated values by
\citet{asplund2009} using the solar spectrum and by \citet{wilms2000}
for the ISM and, therefore, result in a smaller $N_{\rm H}$ value for
a given amount of total absorption due to the matter in the ISM.  We
showed here in our systematic analysis that a change in the abundance
table results in $\approx30\%$ change in the coefficient of the linear
relation between the hydrogen column density and optical extinction.

As a final caveat, we note that when the hydrogen column density is
measured using the ISM abundances (e.g., from
  \citealt{wilms2000}), the relation we presented here should be used
to predict or compare with the extinction in the optical band.  On the
other hand, if an $N_{\rm H}$ is found using solar abundances
(e.g., from \citealt{angr}), then the relation reported in
\citet{go2009} should be employed for self consistency.

\section{Conclusions}

In this paper, we presented a comprehensive study of interstellar
X-ray extinction using the Chandra SNR archive. We used
standardized procedures and made use of the high energy and spatial
resolution of the dataset to assess the uncertainties in the
measurement of the hydrogen column density from X-ray spectra. In
contrast with earlier work, we modeled the interstellar extinction
using the latest ISM model as well as interstellar abundances of \citet{wilms2000}. 
We also modeled SNR spectra with a variety of thermal,
non-thermal, and mixed models to evaluate the effects of the continuum
models on the measured hydrogen column density. In addition to
assessing the uncertainties in this measurement arising from the range
of models, we explored uncertainties due to the spatial variations
within individual remnants and the different physical conditions of
the remnants such as their compositions, temperatures, and
non-equilibrium regions.

We used a Bayesian statistical analysis tools to determine the
systematic uncertainties in the hydrogen column density for each
remnant. We used the hydrogen column density measurements and their
uncertainties in combination with the measurements of the optical
extinction toward these same remnants to determine the relation
between these quantities. We found a best-fit linear relation
described by $N_{\rm H}\ = (2.87\pm0.12)\times10^{21}\ A_{\rm
  V}~{\rm\ cm}^{-2}$.

A couple of different avenues could lead to further progress in the
determination of the empirical relation between the optical extinction
and the hydrogen column density, as also discussed in \citet{go2009}
and \citet{watson2011}. First, a larger number of high quality optical
or near-infrared spectra of SNRs need be obtained to increase the
sample of precise optical extinction measurements.  This is because,
while the extensive archives of X-ray satellites allow us to determine
the hydrogen column density and its uncertainty accurately,
optical/NIR SNR data, especially for strongly reddened regions in the
Galaxy, are scarce. The second strategy targets different populations
of sources that are bright in both optical and X-ray bands, such as
blazars.  The measurement of the hydrogen column density and the
optical extinction in such a population would provide an independent
measurement of the relation between these two quantities and help
cross check the results.

\acknowledgements

This work was supported in part by Chandra Award
No. AR0-11007X. T.G. has been supported by Istanbul University:
Project numbers 49429 and 57321. F.\ \"O. gratefully acknowledges
partial support from NSF grant AST 1108753. P.S. acknowledges partial
support from NASA contract NAS8-03060. This research has made
  use of data obtained from the Chandra Data Archive and the Chandra
  Source Catalog, and software provided by the Chandra X-ray Center
  (CXC) in the application packages CIAO, ChIPS, and Sherpa. The authors thank K. Arnaud for helpful discussions
regarding the \textit{tbabs} model in \textit{xspec}.

\nocite{*}
\bibliographystyle{apj}
\bibliography{bib_drf}

\end{document}